\documentclass[10pt, conference]{IEEEtran}
\usepackage[sort&compress,square,numbers,comma]{natbib}

\ifCLASSINFOpdf
\else
\fi
\usepackage{amsmath,amssymb,amsthm}
\usepackage{graphicx}
\usepackage{bm}
\usepackage{multirow}
\usepackage{subfigure}
\usepackage{color}
\usepackage{url}
\usepackage[english]{babel}
\usepackage{booktabs}
\usepackage{threeparttable}

\begin{document}
%
\title{CASCONet: A Conference dataset}



\author{Dixin Luo, Kelly Lyons\\
Faculty of Information, University of Toronto, Toronto, ON, Canada\\
{\tt\small $\{$dixin.luo,kelly.lyons$\}$@utoronto.ca}
}


%


\maketitle

\begin{abstract}
Knowledge mobilization and translation describes the process of moving knowledge from research and development (R\&D) labs into environments where it can be put to use.
There is increasing interest in understanding mechanisms for knowledge mobilization, specifically with respect to academia and industry collaborations. 
These mechanisms include funding programs, research centers, and conferences, among others.
In this paper, we focus on one specific knowledge mobilization mechanism, the CASCON conference, the annual conference of the IBM Centre for Advanced Studies (CAS). 
The mandate of CAS when it was established in 1990 was to foster collaborative work between the IBM Toronto Lab and university researchers from around the world.
The first CAS Conference (CASCON) was held one year after CAS was formed in 1991. 
The focus of this annual conference was, and continues to be, bringing together academic researchers, industry practitioners, and technology users in a forum for sharing ideas and showcasing the results of the CAS collaborative work.
We collected data about CASCON for the past 25 years including information about papers, technology showcase demos, workshops, and keynote presentations. 
The resulting dataset, called ``CASCONet''\footnote{published on GitHub at \url{https://github.com/iDBKMTI/CASCONet}} is available for analysis and integration with related datasets.
Using CASCONet, we analyzed interactions between R\&D topics and changes in those topics over time. 
Results of our analysis show how the domain of knowledge being mobilized through CAS had evolved over time. 
By making CASCONet available to others, we hope that the data can be used in additional ways to understand knowledge mobilization and translation in this unique context. 
\end{abstract}

\begin{IEEEkeywords}
knowledge mobilization; knowledge translation; CASCON; CASCONet; computer science and engineering; topic models; time series analysis;
\end{IEEEkeywords}

\IEEEpeerreviewmaketitle

\section{Introduction}
There is increasing interest in understanding how knowledge transfer and mobilization takes place. 
At the same time, the number of available datasets and accessible analysis tools is growing. 
Many efforts have been made to make conference datasets available and new techniques have been developed for analyzing conference data for the purpose of understanding outcomes such as knowledge mobilization. 
Vasilescu et al.~\cite{vasilescu2013historical} present a dataset of software engineering conferences that contains historical data about the publications and the composition of program committees for eleven well-established conferences. This historical data is intended to assist conference steering committees or program committee chairs in assessing their selection process or to help prospective authors decide on conferences to which they should submit their work .
Hayat and Lyons analyzed the social structure of the CASCON conference paper co-authorship network and proposed potential actions that might be taken to further develop the CASCON community \cite{hayat2010evolution}. 
They also analyzed the co-authorship ego networks of the ten most central authors in twenty-four years of papers published in the proceedings of CASCON using social network analysis and proposed a typology that differentiates three styles of co-authorship \cite{hayat2017typology}. 
Solomon presented an in-depth analysis of past and present publishing practices in academic computer science conference and journal publications (from DBLP) to suggest the establishment of a more consistent publishing standard  \cite{solomon2009programmers}. 
Many other datasets about conference and journal publications have also been proposed, e.g., the NIPS dataset~\cite{perrone2016poisson}, the Microsoft Academic Graph (MAG)~\cite{sinha2015overview}, and the AMiner database\footnote{\url{https://aminer.org/}}.
Interesting analyses have been proposed and carried out on some of these data sets. 
For example, a relatively new topic model is proposed in~\cite{perrone2016poisson} and verified on the NIPS dataset --- the dynamics of topics on NIPS over time are analyzed quantitatively, e.g., standard neural networks (``NNs backpropagation'') were extremely popular until the early 90s; however, after this, papers on this topic went through a steady decline, only to increase in popularity later on. 
Moreover, the popularity of deep architectures and convolutional neural networks (``deep learning'') steadily increased over these 29 years, to the point that deep learning was the most popular among all topics in NIPS in 2015.
A heterogeneous entity graph of publications is proposed in MAG~\cite{sinha2015overview}, which has potential to improve academic information retrieval and recommendation systems.


In this paper, we describe a specific conference dataset and demonstrate how analyses performed on that dataset can provide insights into mechanisms of knowledge transfer.
We consider the CASCON conference, the annual conference of the IBM Centre for Advanced Studies (CAS). 
The mandate of CAS when it was established in $1990$ was to foster collaborative work between IBM Toronto and university researchers from around the world \cite{perelgut1997overview}.
It is a unique knowledge mobilization and translation environment, specifically designed to facilitate the transfer of technology from university research into IBM products and processes.
The CASCONet dataset presented in this paper is unique in that it includes not only data about authors and papers but also data about all aspects of the CASCON conference. 

\begin{figure*}[!h]
\centering
\includegraphics[width=0.9\linewidth]{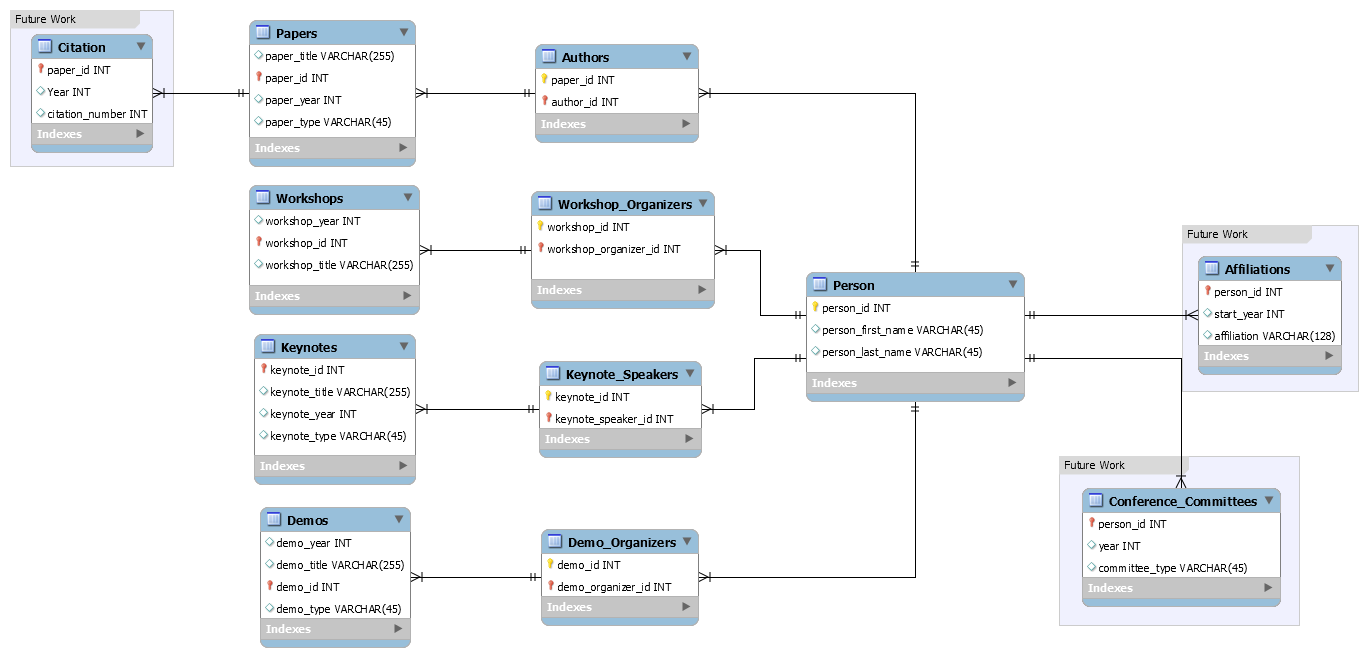}
\caption{
The CASCONet schema
}\label{fig:schema}
\vspace{-.2cm}
\end{figure*}

The first CAS Conference (CASCON) was held in $1991$, one year after CAS was formed. The focus of this conference was, and continues to be, bringing together researchers, government employees, industry practitioners, and technology users in a forum for sharing ideas and results of CAS collaborative work \cite{perelgut1997overview}. 
The CASCON conference is an interesting object of study in this way because it is an annual conference of the IBM Center for Advanced Studies (CAS), a unique center for knowledge mobilization and translation. 
Furthermore, rather than focusing on a narrow topic area in computer science research, CASCON’s mandate is broader, covering many topics in computer science and software engineering with a focus around industry / university collaborations. 
It is therefore interesting to understand what kinds of unique knowledge mobilization structures can be identified by analyzing data about CASCON.

The central data element of the CASCONet dataset is ``person'' and each person's role in CASCON activities is described through the data. 
CASCONet includes the author role and provides title, author, and publication year for over 800 CASCON papers. 
The workshop chair role includes workshop title, workshop chair, and year. 
The keynote data associates people (presenters) with keynote titles and year. 
Finally, the demos data links demo presenter to title and year. 
The people, papers, themes of the workshops, topics of the keynote presentations, and the products and tools presented in the demos over the past 25 years reflect the evolving processes and knowledge mobilization in the CASCON community and may provide a glimpse into the field of computer science (advanced methods and techniques, challenges and urgent problems, innovation, and applications) overtime. 
As an example of the kinds of analyses that can be performed on this dataset, we present basic statistics of CASCON and analyze the temporal dynamics of topics presented at CASCON.
We believe that this dataset and analyses such as these may provide researchers of computer science and social science with a new resource to study the co-evolution of academic and industry communities. 

\section{Properties of CASCONet}
The first CASCON took place in $1991$. 
More than $1500$ researchers, technologists, developers, and decision makers attend CASCON each year.
CASCONet $(1991 - 2016)$ contains data about a total of $2517$ people who have written $846$ papers, presented $1212$ demos, delivered $107$ keynote presentations, and organized $796$ workshops. Fig.~\ref{fig:schema} shows the schema of CASCONet.

\begin{figure}[!ht]
\center
\subfigure[Numbers of papers]{
\includegraphics[width=0.95\columnwidth, height=3.5cm]{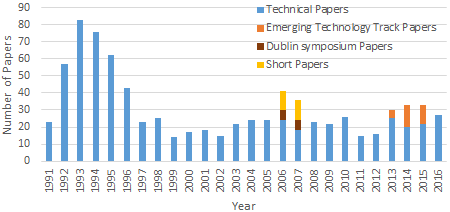}\label{fig:paperyear}
}
\subfigure[Numbers of workshops]{
\includegraphics[width=0.95\columnwidth, height=3.5cm]{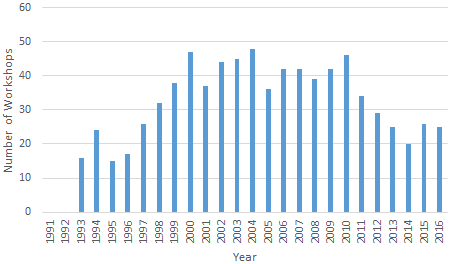}\label{fig:workshopyear}
}
\subfigure[Numbers of demos]{
\includegraphics[width=0.95\columnwidth, height=3.5cm]{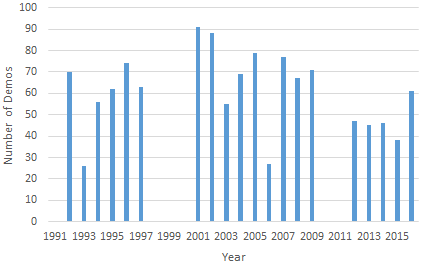}\label{fig:demoyear}
}
\caption{
The basic statistics of CASCON over time
}
\vspace{-.2cm}
\end{figure}

\begin{table*}[!t]
\centering
\caption{Top $10$ words of $10$ topics}\label{table:topwords}
\vspace{-8pt}
\begin{threeparttable}[c]
\begin{tabular}{ l | l}
\hline\hline
Topic & Top 10 words \\ \hline\hline
Users and Business & User, Internet, task, trust, model, resource, web, information, database, search \\ \hline
 Cloud and Web Services & Cloud, web, service, application, design, development, integral, user, code, URL \\ \hline
 Systems & System, model, distributed, computing, database, user, management, paper, application, information \\ \hline
 Programs and Code & Program, performance, compiler, language, parallel, class, oriented, object, code, Java \\ \hline
 Applications & Interaction, application, user, mobile, interface, device, information, visual, tool, support \\ \hline
 Networks and Security & Security, network, cache, local, communication, enterprise, layer, server, grid, privacy\\ \hline
 Software & Software, design, analysis, engine, tool, approach, test, development, process, performance \\ \hline
 Databases & Database, usage, system, transaction, optimization, query, DB2, data, user, system \\ \hline
 Data Analysis & Data, analytic, user, mining, decision, event, distribution, information, business, system \\ \hline
 Algorithms & Algorithm, problem, performance, architecture, system, cluster, design, time, schedule, test \\ \hline\hline
\end{tabular}
\end{threeparttable}
\end{table*}

{\bf Person.}
According to the dataset, $24.0\%$ of the people who authored papers at CASCON have published more than one paper in CASCON.
The most number of papers published by one person is $20$. 
There are $33$ people whose time span of authoring papers in CASCON is greater than $10$ years. 
There are only $4$ people who have participated in CASCON in all roles, as author, workshop chair, demo organizer, and keynote speaker.

{\bf Papers.}
Fig.~\ref{fig:paperyear} shows the number of CASCON papers published each year.
The CASCON main conference has accepted a relatively stable number of papers for each year since $1997$.
In the early years, a greater number of papers were accepted to CASCON. Then in $2006$ and $2007$, in addition to the main conference, an IBM Dublin CAS symposium was held. An ``emerging technology track'' and a ``short paper track'' were added in $2013$. 
In CASCONet, the papers are identified by their types: ``technical papers''; ``short papers''; ``emerging papers''; or, ``symposium papers''.

{\bf Workshops and demos.} Fig.~\ref{fig:workshopyear} illustrates the number of workshops held at CASCON each year. 
Workshops provide attendees with opportunities to learn new technologies, learn about concepts, identify collaboration opportunities, share results, and engage in discussions around specific topics.
The dynamics of the number of workshops is quite different from that of the number of papers published each year. The years between $2000$ and $2010$ saw a greater number of workshops than previous to then or since.
Fig.~\ref{fig:demoyear} shows the numbers of demos over the years.
Note that there are not any records available for demos exhibited between $1998$-$2000$ and for $2010$ and $2011$.


\section{Topics of CASCONet Papers Over Time}
In this section, we establish a topic model for CASCON papers, with the aim to capture the evolution of topics over time, and quantify and analyze the influences between pairs of topics. This is one of many kinds of analyses that can be carried out on the CASCONet dataset.

{\bf Topic Model of Papers.} The topic model of CASCON papers was extracted using Latent Dirichlet allocation (LDA)~\cite{blei2003latent}.
We collected the nouns, verbs, and adjectives in the titles of all the $M=846$ CASCON papers from $1991$ to $2016$, and built a word corpus.
$N$ ``topics'' were extracted  as bags of words using LDA, by calculating the conditional probabilities $[p(\mbox{word}|\mbox{topic})]$ and $[p(\mbox{topic}|\mbox{paper})]$.
We first set $N=50$ and learned the LDA model. 
Then, using $[p(\mbox{word}|\mbox{topic})]$ as features of topics, we clustered topics into $N=10$ higher-level topics based on the correlations among the topic features. 
The $10$ words having the highest conditional probability $p(\mbox{word}|\mbox{topic})$ in each topic are shown in Table~\ref{table:topwords}. 
We manually summarized the semantic meaning of topics and labeled them as ``users and business'', ``cloud and web services'', ``systems'', ``programs and code'', ``applications'', ``networks and security'', ``software'', ``databases'', ``data analysis'' and ``algorithms''.  

\begin{figure}[!ht]
\center
\subfigure[Topic distribution over time]{
\includegraphics[width=0.95\columnwidth, height=4.5cm]{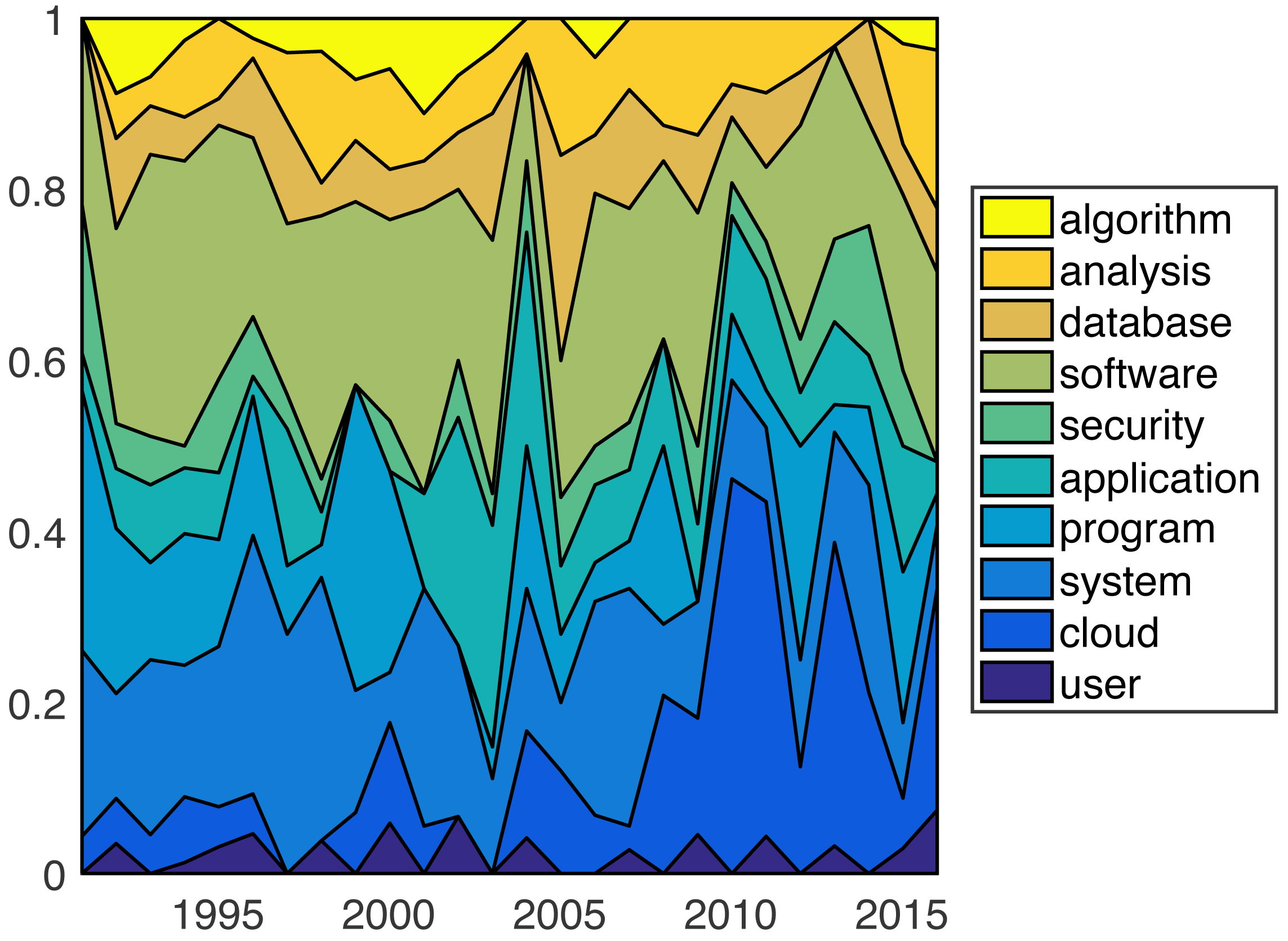}\label{fig:dis}
}
\subfigure[Box plot of topic distribution]{
\includegraphics[width=0.95\columnwidth, height=4.5cm]{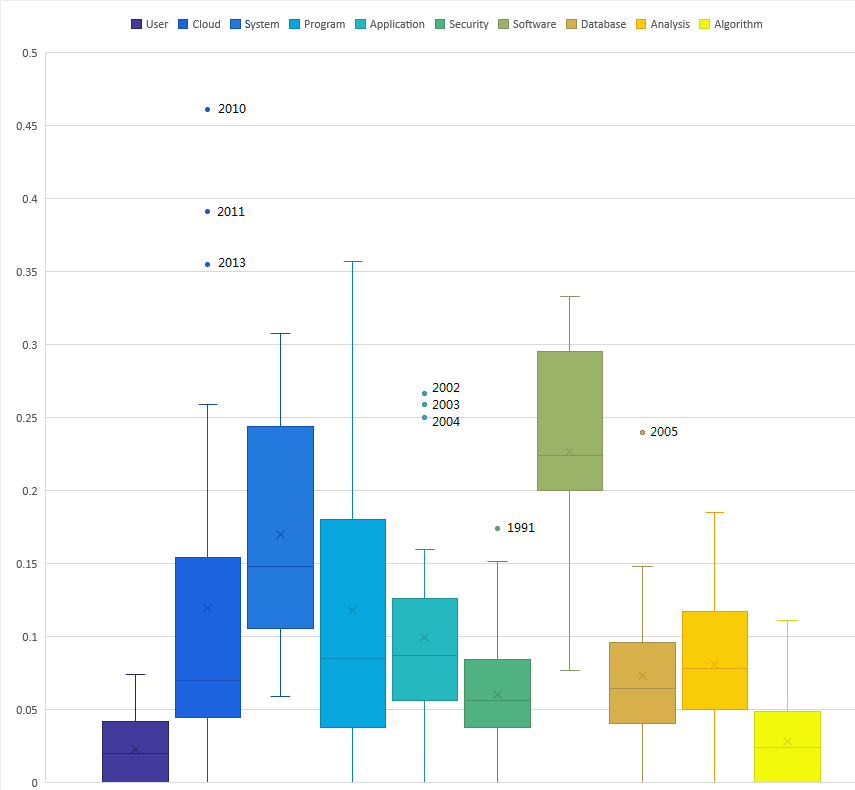}\label{fig:disbox}
}
\caption{
The visualization of topic distribution.
}\label{fig:topdis}
\vspace{-.2cm}
\end{figure}

For each paper, we can take the the topic with the highest conditional probability as the theme of the paper. 
We can then calculate the distribution of topics over years by counting the number of papers corresponding to different topics in each year. 
Fig.~\ref{fig:topdis} visualizes the distribution of topics over year. 
In the early years of CASCON ($1991$-$1995$), we see that ``software'', ``systems'', and ``programs and code'' are the predominant topics ---  papers about these three topics occupy most of the accepted submissions.
Between 1991 and 1995, the average percentage of CASCON papers about ``software'', ``systems'', and ``programs and code'' are $28.1\%$, $17.7\%$, and $17.8\%$, respectively. 
With the development of interest in cloud computing~\cite{armbrust2010view}, more papers in this area appeared in CASCON starting in $2008$.
Moreover, ``software'' is one of the main topics in nearly all of the $26$ CASCON conferences, while the numbers of papers in ``users and business'', ``networks and security'' and ``algorithms'' are relatively small but stable over all years.

{\bf Granger Causality Analysis of Topics.} After identifying the topics hidden in the papers' titles, we further analyzed the causal relationships among the topics. 
We developed a dynamic model of topics based on a vector auto-regressive (VAR) model~\cite{lutkepohl2011vector} and analyzed the Granger causality~\cite{granger1969investigating,basu2015network,xu2016learning} of topics based on this model.
Specifically, we calculated the distributions of the topics over the years according to the results of LDA model, denoted as $[\bm{p}_1, ..., \bm{p}_T]\in\mathbb{R}^{N\times T}$. 
Considering $\{\bm{p}_t\}_{t=1}^T$ as an instance of time series, we describe its transition process by learning the following 1-order vector auto-regressive model:
\begin{eqnarray}\label{var}
\begin{aligned}
\min_{\bm{A}}~\sum_{t=1}^{T-1}\|\bm{p}_{t+1}-\bm{A}\bm{p}_{t}\|_2^2 +\lambda\|\bm{A}\|_1.
\end{aligned}
\end{eqnarray}
Here $\bm{A}=[a_{ij}]$ is a transition matrix, whose element $a_{ij}$ measures the influence $j$-th topic imposes on the $i$-th topic. 
If $a_{ij}>0$ ($<0$), the $j$-th topic is understood to trigger (suppress) the $i$-th topic in the next time-stamp (in this case, year). 
$a_{ij}=0$ if, and only if, the $i$-th topic is locally independent on the $j$-th topic. 
The first term of the objective function minimizes the estimation error of the time series. 
Furthermore, considering the fact that generally a topic is only related to a subset of topics~\cite{luo2016learning,luo2015multi}, we impose a sparse constraint, $\|\bm{A}\|_1=\sum_{i,j}|a_{ij}|$, on the transition matrix $\bm{A}$. 
The VAR model can be learned effectively by using the alternating direction method of multipliers (ADMM) method. 

\begin{figure}[!h]
\centering
\includegraphics[width=0.7\linewidth]{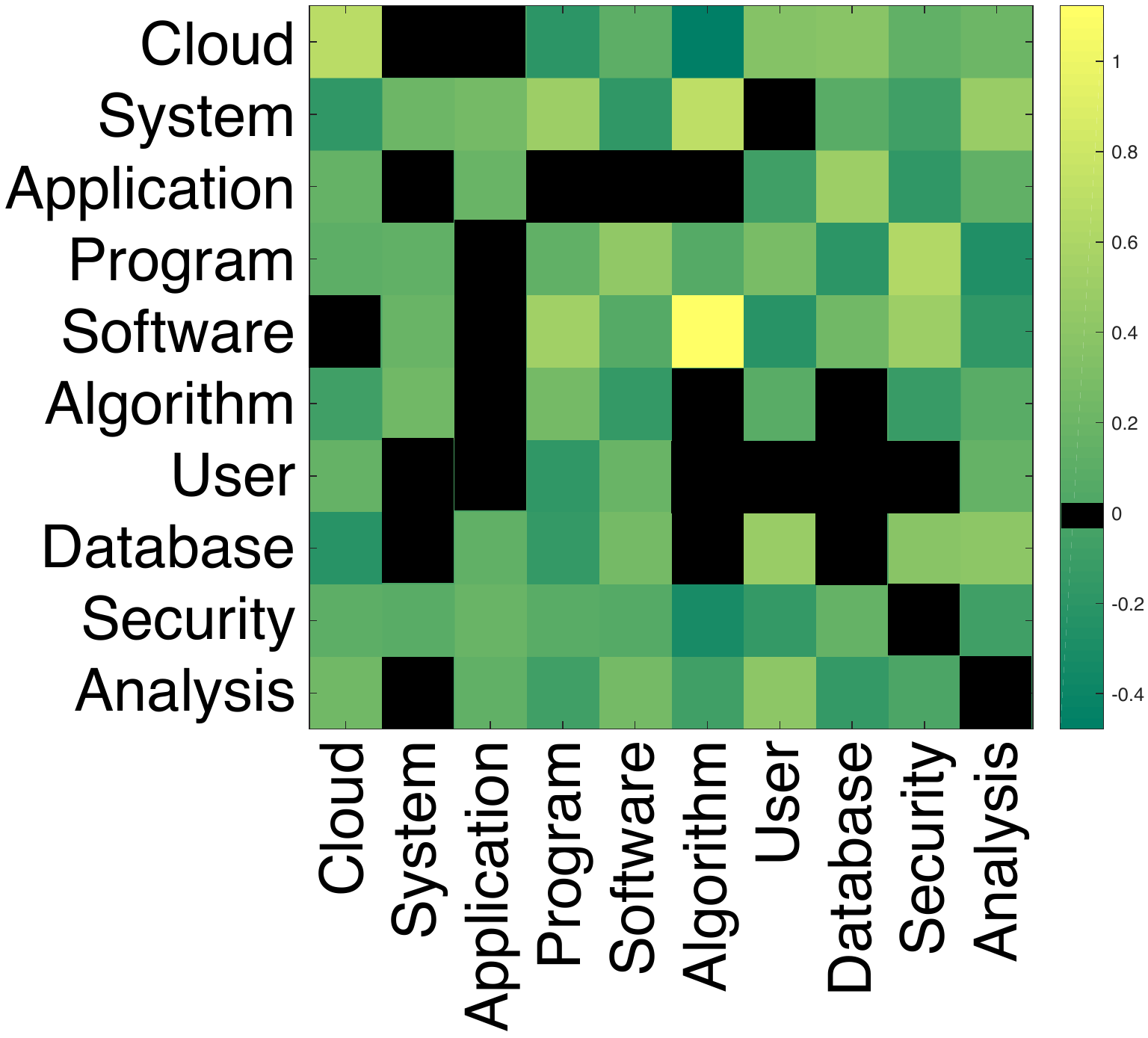}
\caption{The transition matrix of topics.}\label{fig:TransMat}
\end{figure}

According to the work in~\cite{basu2015network,xu2016learning}, the transition matrix is actually the adjacent matrix of a Granger causality graph $G(\mathcal{N},\mathcal{E})$~\cite{granger1969investigating}, where the node set $\mathcal{N}=\{1,...,N\}$ contains topics and the edge set $\mathcal{E}$ indicates the Granger causality of topics. 
For $i,j\in\mathcal{N}$, we say that the $j$-th topic ``Granger causes'' the $i$-th topic if and only if $j\rightarrow i\in \mathcal{E}$, which is equivalent to $a_{ij}\neq 0$. 
Fig.~\ref{fig:TransMat} shows the inferred transition matrix, where the topics are sorted in descending order according to their self-triggering intensity $a_{ii}$, $i=1,...,10$. 
We find several interesting phenomena.

\begin{itemize}
\item \textbf{Endogenous and exogenous topics.} In Fig.~\ref{fig:TransMat}, the element $a_{ij}$ is labeled as black when $|a_{ij}|<0.02$, which indicates that the triggering from $j$ to $i$ is so weak that we can ignore the Granger causality from $j$ to $i$. 
We find that ``cloud and web services'', ``systems'', ``applications'', ``programs and code'' and ``software'' have positive $a_{ii}$ while the other five topics have $a_{ii}=0$. 
This means that these five topics (``cloud and web services'', ``systems'', ``applications'', ``programs and code'' and ``software'') are endogenous topics of CASCON, which, once introduced, tend to appear continuously over the years, while the other five topics are exogenous topics influenced by endogenous topics. 
In other words, for each exogenous topic, its appearance is mainly caused by other topics rather than itself, and its appearance in any given year does not contribute to its future appearance. 
From this perspective, it seems that CASCON is a conference focused on the fundamental engineering problems of computer science, e.g., systems and software engineering. 
The higher-level problems, such as user experience, network security, data analysis and so on, seem to be more complementary to the conference. This observation may relate to the focus of the IBM Toronto Lab and knowledge mobilization within the CAS community.
\item \textbf{Positive and negative triggering patterns.} 
We do not impose any nonnegative constraints on the optimization problem~(\ref{var}) so that some of the elements of $\bm{A}$ are negative. 
We find that most of the $a_{ij}$'s are positive (light green and yellow in Fig.~\ref{fig:TransMat}) while some of them are negative (dark green ones in Fig.~\ref{fig:TransMat}). 
Positive triggering patterns indicate that the corresponding two topics are highly-correlated with each other. 
For example, the ``Security'' topic is likely triggered by other topics because the security problem is related to many fields of computer science, e.g., system security, cloud security, client information security, and security algorithms, etc. 
On the other hand, negative suppressing patterns may indicate that the corresponding topics have a competitive relationship in CASCON. 
It is natural that when the time available for paper presentations at a conference is fixed, the total number of publications is limited as well. 
As a result, when the number of papers about one topic is dominant at the conference, the submission and acceptance of paper on other topices may be suppressed accordingly. 
For example, the ``systems'' topic is a technical topic while the ``users and business'' topic is a business-related topic. 
These two topics are competitive --- papers on systems tend to be selected over papers on business topics and vice versa. 
The inverse phenomenon would happen when organizers want to increase business interactions.
\end{itemize}

\section{Conclusion}
CASCON is a longtime conference held by the IBM Canada Lab Centre for Advanced Studies (CAS), which provides us with an interesting dataset ``CASCONet''. 
In this paper, we introduced the data we collected and presented some preliminary analyses. 
Our CASCONet dataset is available on \url{https://github.com/iDBKMTI/CASCONet}. 
In the future, we will collect more data about IBM CAS, and study the correlation between CASCON and the development of CAS itself, in order to better understand the mechanism of collaboration, knowledge translation and output. 
\section*{Acknowledgment}
This research was partially funded by an NSERC Strategic Partnership Grant.

\bibliographystyle{IEEEtran}
\bibliography{bare_conf}

\end{document}